\documentclass[journal]{IEEEtran}

\ifCLASSINFOpdf
\else
\fi

\IEEEoverridecommandlockouts
\usepackage{amsmath,amssymb,amsfonts}
\DeclareMathOperator*{\argmax}{argmax}
\usepackage{algorithmic}
\usepackage{fancyhdr, graphicx}
\usepackage{textcomp}
\usepackage{xcolor, colortbl}
\usepackage[utf8]{inputenc}
\usepackage[english]{babel}
\usepackage[autostyle]{csquotes}
\usepackage[ruled,vlined]{algorithm2e}
\usepackage{csvsimple}
\usepackage{bbm}
\usepackage{caption}
\usepackage{subcaption}
\usepackage{comment}

\definecolor{Gray}{gray}{0.8}
\definecolor{Gray2}{gray}{0.9}

\begin{document}

\title{Classification of Radio Signals Using Truncated Gaussian Discriminant Analysis of Convolutional Neural Network-Derived Features}

\author{J.B. Persons,~\IEEEmembership{Graduate Student Member,~IEEE,}
        L. J. Wong,
        \\W. C. Headley,~\IEEEmembership{Member,~IEEE,}
        and~M. C. Fowler,~\IEEEmembership{Member,~IEEE}
\thanks{J.B. Persons is a Ph.D. student of the Department
of Electrical and Computer Engineering, Virginia Tech, Blacksburg,
VA, 24061 USA e-mail: persons@vt.edu.}
\thanks{Lauren Wong is a deep learning data scientist with Intel Corporation, San Jose, CA, USA.}
\thanks{William ``Chris" Headley and Michael Fowler are research faculty with the Hume Center for National Security and Technology, Virginia Tech.}
}

\markboth{Under peer review for IEEE Journal on Selected Areas in Communications, January 2021}%
{Persons \MakeLowercase{\textit{et al.}}: Classification of Radio Signals Using Truncated Gaussian Discriminant Analysis of Convolutional Neural Network-Derived Features}

\maketitle

%
\IEEEpeerreviewmaketitle

\begin{abstract}
To improve the utility and scalability of distributed radio frequency (RF) sensor and communication networks, reduce the need for convolutional neural network (CNN) retraining, and efficiently share learned information about signals, we examined a supervised bootstrapping approach for RF modulation classification. We show that CNN-bootstrapped features of new and existing modulation classes can be considered as mixtures of truncated Gaussian distributions, allowing for maximum-likelihood-based classification of new classes without retraining the network. In this work, the authors observed classification performance using maximum likelihood estimation of CNN-bootstrapped features to be comparable to that of a CNN trained on all classes, even for those classes on which the bootstrapping CNN was not trained. This performance was achieved while reducing the number of parameters needed for new class definition from over 8 million to only 200. Furthermore, some physical features of interest, not directly labeled during training, e.g. signal-to-noise ratio (SNR), can be learned or estimated from these same CNN-derived features. Finally, we show that SNR estimation accuracy is highest when classification accuracy is lowest and therefore can be used to calibrate a confidence in the classification.
\end{abstract}

\begin{IEEEkeywords}
radio frequency machine learning (RFML), supervised bootstrapping, truncated Gaussian discriminant analysis (TGDA), maximum likelihood estimation (MLE)
\end{IEEEkeywords}



\section{Introduction}
\subsection{Motivation}
\IEEEPARstart{T}{he} proliferation of wireless communication devices and increasing complexity of the radio frequency (RF) environment are creating unprecedented strains on the ability of communication networks to maintain requisite situational awareness. \textit{Ex post} shared spectrum approaches, for example, require awareness of participating transmitters for deconfliction and identification of non-compliant actors \cite{bhattarai_overview_2016}, a task exacerbated as the number of transmitters increases. To reduce the resource demands of establishing and maintaining large RF communication or sensor networks, it is necessary to keep down the size, weight, and power requirements, as well as cost (SWAP-C) of each network node while minimizing collaborative transmission requirements. For the purposes of signal classification, this may mean utilizing automated classification methods that do not require excessive pre- or post-processing and that can be updated with new classes of interest without large-scale data transfer.  An automated classification system should also be able to integrate additional information that was not available or relevant at the time of network establishment and training into its classification decisions.

The goal of the research outlined in this paper is to investigate one promising approach to signal classification that would reduce the need for retraining neural networks when adding classes, allow classes of interest to be concisely characterized, and facilitate the incorporation of additional data sources into classification decisions.

\subsection{Overview}
Adapting a CNN approach previously used for signal classification \cite{semisup}, we started with the assumption that classes could be represented by mixtures of multivariate Gaussian distributions across the learned feature set. We initially attempted unsupervised clustering of BPSK, QPSK, QAM16, and QAM64 signals using intermediate layer outputs of a CNN as features, with the intent of learning mixture parameters for inference; however, the clustering algorithm would consistently subdivide the BPSK and QPSK signal classes many times before finding any meaningful distinction between the QAM signal classes. The lack of useful clustering in this approach led us to turn to a supervised method instead. 

While our initial intuition that classes could be represented by Gaussian mixtures was partially validated, we discovered that mixture components were better modeled as mixtures of point masses and truncated Gaussians. Using truncated Gaussian discriminant analysis (TGDA) , we drew empirical distributions of features from labeled sets, then conducted maximum likelihood estimation (MLE) for each class.  Despite being trained only to infer modulation classes, we discovered that the CNN-derived features used for modulation classification could also be used for signal-to-noise ratio (SNR) and samples-per-symbol (SPS) estimation. We repeated this process using networks trained on different types and numbers of training classes in order to examine the effects of the base network on TGDA performance.

In accordance with our goals, this approach offers several advantages over a purely CNN-based system. Once sensors are equipped with the trained neural network, new class descriptions can be shared among the classes using only 200 parameters, rather than the several million that would be required to characterize a retrained neural network. By inputting signals with known physical characteristics into the feature extraction network, we can learn the network feature representation of those physical features. As we are using MLE for classification, we inherently have a measure of confidence built into our approach. Because the classification algorithm uses the neural network only to generate features, there is an opportunity to include other information as additional features the MLE calculation, though we do not explore that possibility in this work.

We discuss our training and inference approach in Section \ref{Technical Approach}, compare the performance of this approach to a baseline CNN in Section \ref{Experiments}, and assess how changing training data classes affects the classification performance and generalizability of TGDA MLE in Section \ref{Comparative Experiments}.

\section{Related Work}
Previous research in radio frequency machine learning (RFML) \cite{OShea} demonstrated the efficacy of using convolutional neural networks (CNNs) to determine the modulation classification of a signal based on received in-phase and quadrature (I/Q) data. Follow-on work applied the concept of supervised bootstrapping, in which the final (softmax) layer of a trained neural network classifier is discarded, and the remainder of the network is used for feature extraction \cite{semisup, Clustering}. These works then applied unsupervised clustering algorithms to the features derived from the bootstrapping approach, demonstrating the feasibility of identifying new classes from unlabeled data.  Wong, \textit{et al.} also observed indications that CNNs learn different features from different training signal bandwidths, a fact that will be expanded upon within this work \cite{Clustering}. Later research showed that CNNs could use I/Q data for specific emitter identification or separation of interfering signals \cite{wong_specific_2019, shi_deep_2019} and measured the effects of frequency and sample rate offsets on CNN performance \cite{hauser_signal_2017}.

Subsequent RFML approaches primarily have been aimed at improving classification accuracy. Karra, \textit{et al.} used multiple neural networks in a hierarchical approach to determine data type, modulation class, and modulation order \cite{karra_modulation_2017}. This work also noted the limitations imposed by small ``snapshot" size - what we will refer to in this work as signal snippets, or the I/Q input to the neural network - and recommended using longer snapshots when available for better feature extraction. In \cite{mossad_deep_2019}, the authors employed multiple CNNs with both I/Q and frequency domain (fast Fourier transform (FFT)) inputs in a multitask learning scheme to improve classification accuracy in the modulation classes that showed higher levels of confusion in \cite{OShea}. The authors of \cite{zeng_spectrum_2019} applied CNNs in a more conventional sense by conducting spectral analysis with FFTs, then feeding the spectrogram image into a CNN for modulation classification, achieving higher classification accuracies than in \cite{OShea} at the cost of higher memory and training time requirements. Though Youssef \textit{et al.}'s focus was the introduction of a multi-stage training approach, the authors also noted the improved performance of CNNs relative to conventional deep neural networks as input size increased \cite{youssef}.

As with the original RFML work, some insight may be gained from techniques being applied in the computer vision field. The authors of \cite{sun_deep_2014} used 60 CNNs to extract visual features, represented these features as sums of Gaussian variables, and then performed facial verification by the Joint Bayesian technique. Other computer vision work fused CNN-derived features with ''traditional" features from medical imaging, then used these new features as inputs into a multi-layer perceptron classifier to achiever higher levels of accuracy than the CNN could achieve alone \cite{lai_medical_2018}. Encouragingly for our work, Donahue, \textit{et al.} demonstrated that the later layers of neural networks were able to capture the high level features required for both semantic classification and subclass recognition \cite{decaf}.

\section{Technical Approach}\label{Technical Approach}

\subsection{Data Generation}
To obtain signals for training, validation, and testing, we first generated a random bitstream, then used GNU Radio to create 2048-sample snippets of I/Q data of desired signal types, and each snippet was passed through a root-raised cosine filter with a roll off factor of 0.35, without loss of generality. Additive white Gaussian noise (AWGN) was added to achieve the desired signal-to-noise ratio (SNR). For simplicity in exploring this approach to signal classification, no frequency or sample rate offsets were imposed.

\subsection{Neural Network for Feature Extraction}
\begin{figure}
    \centering
    \includegraphics[width=\columnwidth]{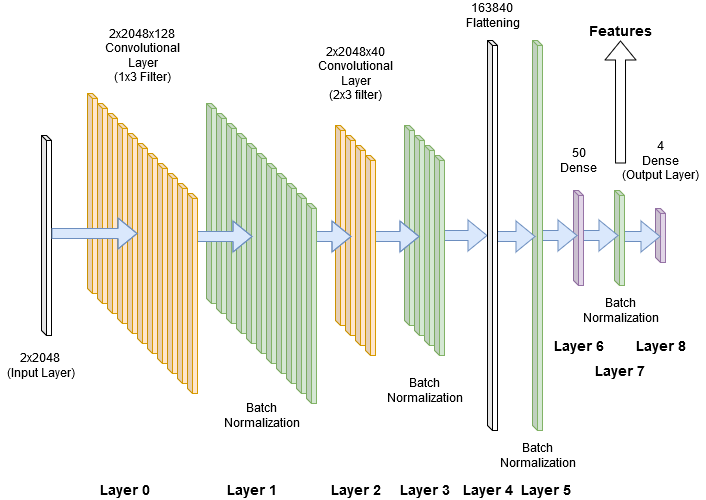}
    \caption{Structure of CNN used for classification of original modulation classes, as defined in Keras. Features used for distribution-based classification are drawn from the outputs of the batch-normalized dense layer.}
    \label{fig:structure}
\end{figure}

The CNN architecture used for feature extraction in this work is shown in Figure \ref{fig:structure}, and is based loosely off of that given in \cite{OShea}.
However, the overall size of the network was reduced to accommodate a simpler classification problem with fewer output classes (and be trained on an older GPU with limited memory), and batch normalization was used in place of dropout as the already reduced layer size did not offer much tolerance for additional sparsity.
Additionally, input size was increased to allow a larger number of samples of I/Q data to give the network more information with which to make decisions, and the hyperbolic tangent activation function was used instead of the ReLU activation function following the two convolutional layers to encourage a smoother distribution of features and allow for negative outputs from each layer. Though this network is less complicated than state-of-the-art image networks at 8,551,642 trainable parameters, it is sufficiently robust to achieve test accuracy of greater than 98 percent across all four original training classes at 20 dB SNR.  Additionally, its relatively small memory requirement makes it a better surrogate for the type of neural network that might see deployment on low-SWAP-C sensors.

In this work, five networks were trained for comparison purposes using this architecture in order to stress the generalizabilty of the develop approach.  Each network was trained on 16,000 2048-sample I/Q signal snippets of each modulation class at random float-valued SNRs from a uniform distribution of range 0 to 20 dB, then validated on 4,000 signal snippets of each class pulled from the same SNR distribution.  The training classes in each of the five models are depicted in Table \ref{tab:training_networks}.
\begin{table}
    \centering
    \begin{tabular}{|c|c|}
    \hline
         Model 1 & BPSK, QPSK, QAM16, QAM64 at 4 sps\\ \hline
         Model 2 & BPSK, QAM16, QAM64, GFSK at 4 sps\\ \hline
         Model 3 & BPSK, QPSK, QAM16, QAM64, GFSK at 4 sps\\ \hline
         Model 4 & BPSK, QPSK, QAM16, QAM64, GFSK at 2, 4, and 8 sps\\ \hline
         Model 5 & BPSK, QPSK, QAM16, QAM64, VT, GFSK, GMSK at \\
         & 2, 4, and 8 sps\\
    \hline
    \end{tabular}
    \caption{Model names and training classes used in this work.}
    \label{tab:training_networks}
\end{table}

\subsection{Feature Distributions}
\begin{figure}[!t]     
    \begin{center}
    \fbox{\includegraphics[width=1.5in]{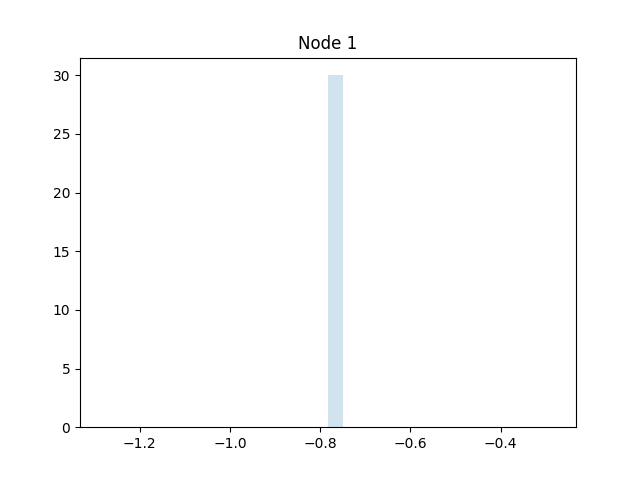}}   
    \subcaption{Probability point mass ``spike."}
    \vspace{0.2in}
    \fbox{\includegraphics[width=1.5in]{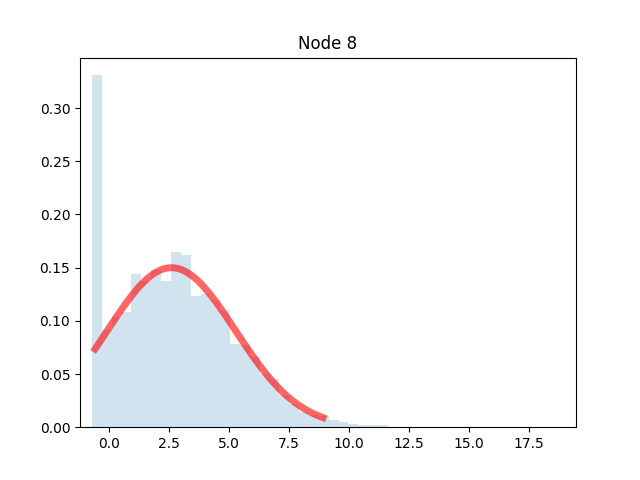}}
    \subcaption{Spike with truncated Gaussian.}
    \vspace{0.2in}
    \fbox{\includegraphics[width=1.5in]{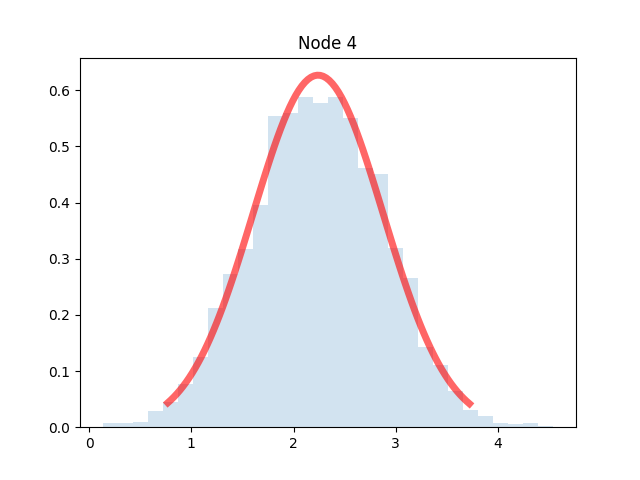}}
    \subcaption{Gaussian distribution.}
    \end{center}
    \caption{Example empirical distributions of modulation class features.}
    \label{DistTypes}
\end{figure}

Despite the network's performance, it still suffers from the inherent limitations of a CNN in that it must be retrained if new classes are added.
We circumvented this weakness of the CNN by first training normally, then discarding the classification layer (layer 8) of the trained network and using outputs from the normalized first dense layer (layer 7) as features for classification, termed supervised bootstrapping, as in \cite{semisup}. It was apparent to the authors that the different modulation classes have distinct feature distributions at 20dB SNR, though these distinctions diminish at lower SNRs - QAM16 and QAM64 distributions are nearly indistinguishable to the eye at 5dB SNR. Also, our assumption that class mixture elements could be represented as multivariate normal distributions (or ``spikes" in the limiting case) appeared reasonable for only about half of the features in a given class mixture element.  Upon closer examination, we found that class features actually appear as mixtures of point masses, Gaussians, and truncated Gaussians as depicted in Figure \ref{DistTypes}. To characterize each class of interest, we fed 5,000 signal snippets of each class through the neural network and extracted four parameters for each of the 50 nodes of layer 7: $s$, the value of the lowest element in the feature (or spike); $p$, the percentage of the probability mass in the spike; and the location and scale parameters of the (truncated) Gaussian. We were able to estimate these parameters efficiently using available software packages due to the helpful observation that the probability spikes, when present, are equal in mass to the truncated portion of the Gaussians for any given feature.

\subsection{Inference Approach}
Once parameters for the feature distributions of each class of interest were determined, inference was performed through maximum likelihood estimation using 

\begin{equation}
\begin{aligned}
    &\text{class}(x) = \argmax_c \sum_{i=1}^{50} \log \left\{ \mathbbm{1}(p_{ci}\geq 0.01)p_{ci}[\mathbbm{1}(x=s_{ci})\right.\\
    &+\mathbbm{1}(p_{ci}\geq 0.01)(1-p_{ci})\text{Trunc}\mathcal{N}(a_{ci}, \text{loc}_{ci}, \text{scale}_{ci})|_{x_{ci}}]\\ 
    &+ \left.\mathbbm{1}(p_{ci}<0.01)\mathcal{N}(\text{loc}_{ci}, \text{scale}_{ci})|_{x_{ci}}\right\}, \label{MLE}\\
\end{aligned}
\end{equation}

\noindent where $a_{ci}=(s_{ci}-\text{loc}_{ci})/\text{scale}_{ci}$ is required to convert clip values from those of a truncated standard normal to the appropriate mean and standard deviation \cite{scipy}. The $0.01$ threshold was established arbitrarily as the minimum percentage of probability mass which may constitute a spike. Using the log likelihood is helpful for numerical stability, and if we wished to include additional features in our classification process, we could do so simply by adding the (weighted) log likelihood of these additional features. Note that for the purposes of minimizing parameters and computation, we assume independence among features in this exploratory work.

\section{Single-Network Experiments}\label{Experiments}
We conducted several experiments to determine the efficacy of our approach.  We first tested performance of truncated Gaussian discriminant analysis against the original network in classifying the four base modulation classes on which the network was trained.  We next attempted classification of modulation classes on which the network had not been trained.  Then, we tested the utility of our approach in determining physical features of the signals for which the network was not designed to classify: signal-to-noise ratio and samples per symbol.

For a performance metric, we elected to use the balanced F measure, or $F_1$ score, as it better captures the interplay between inputs in a multi-class classifier than accuracy alone. While $F_1$ score varies with class size in unbalanced datasets \cite{manning}, we structured our experiments in such a way that all classes were equally represented, avoiding this potential problem. We calculate the $F_1$ score using

\begin{equation}\label{F1}
    F_1 = 2*\frac{precision*recall}{precision+recall},
\end{equation}
where 
\begin{equation}
    precision = \frac{\textit{true positive}}{\textit{true positive} + \textit{false positive}}
\end{equation}
and
\begin{equation}
    recall = \frac{\textit{true positive}}{\textit{true positive} + \textit{false negative}}.
\end{equation}

\subsection{Modulation Classification - Model 1}
\subsubsection{Classification of Trained Classes}\label{Classification of Trained Classes}
To assess baseline performance of TGDA in this application, we calculated mean vectors and covariance matrices for each class using 5,000 signal snippets of each class taken at SNRs varying uniformly from 0 to 20 dB.  We then tested both the baseline network and the TGDA approach against 1000 signal snippets per class at SNRs varying uniformly randomly from 0 to 20dB (4,000 snippets in total).  The results of this assessment can be seen in the 4-Class column of Table \ref{combined_F1_5class}.  For reference, the $F_1$ score for a random classifier in this case would be 0.25.  As the reader can see, the performance of the TGDA approach is comparable to that of the baseline network even without employing mixtures of SNR-specific distributions for each class.

\subsubsection{Classification with Additional Classes - Model 1}
\begin{figure}[t]
\begin{center}
\includegraphics[width= \columnwidth]{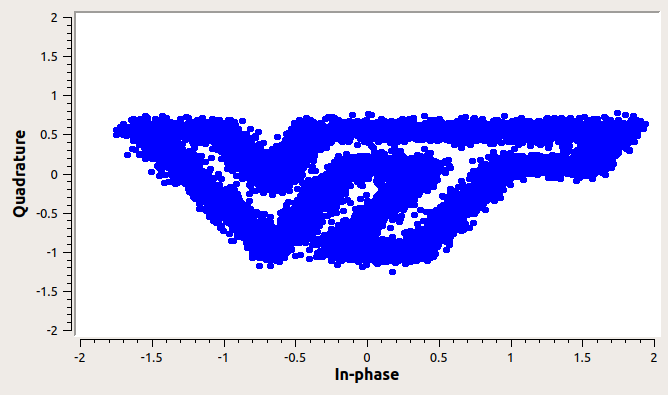}
\end{center}
\captionsetup{width = \columnwidth}
\caption{The 64-VT modulation scheme, while impractical for communications, is a phase- and amplitude-modulated signal similar to those used to train Model 1. Here its I/Q representation is displayed at 20dB SNR, the highest SNR used in these experiments.}
\label{VT}
\end{figure}

We next calculated feature parameters for each class using the same method in Section \ref{Classification of Trained Classes}, except we added a fifth class, the 64-VT modulation, created in \textit{liquid-dsp} and displayed in Figure \ref{VT} \cite{liquid}. The 64-VT modulation, while not practical, is phase- and amplitude-modulated like the training set and has the added benefit of displaying the Virginia Tech logo in the I/Q plane.  The neural network had not been trained on this waveform, so the base network had no means to correctly classify this class.  However, using TGDA, we were able to achieve classification accuracies of the new class comparable to those in the original test without significantly reducing performance among the four original classes - though we did see confusion between the QPSK and the VT signals of 2.5\%. Results are displayed in the 5-Class column in Table \ref{combined_F1_5class}.  For reference, the $F_1$ score for a random classifier in this case would be 0.2.

\begin{table}[t]
\caption{$F_1$ Scores for 4- and 5-Class Classification of Base Network and TGDA with Single Component Per Class in Classifying 1000 Signals Between 0 and 20 dB SNR}
\begin{center}
\renewcommand{\arraystretch}{1.2}
\begin{tabular}{|c|c|c|c|}
\hline
\textbf{Mod Class} & \textbf{Base Network} & \textbf{4-Class} & \textbf{5-Class}\\
 &  & \textbf{TGDA} & \textbf{TGDA}\\ \hline
BPSK & 1.000 & 1.000 & 0.997 \\ \hline
QPSK & 0.929 & 0.949 & 0.917 \\ \hline
QAM16 & 0.680 & 0.703 & 0.674 \\ \hline
QAM64 & 0.715 & 0.716 & 0.691 \\ \hline
VT & N/A & N/A & 0.967 \\ \hline
\end{tabular}
\end{center}
\label{combined_F1_5class}
\end{table}

\begin{figure}[t]
\begin{center}
\includegraphics[width= \columnwidth]{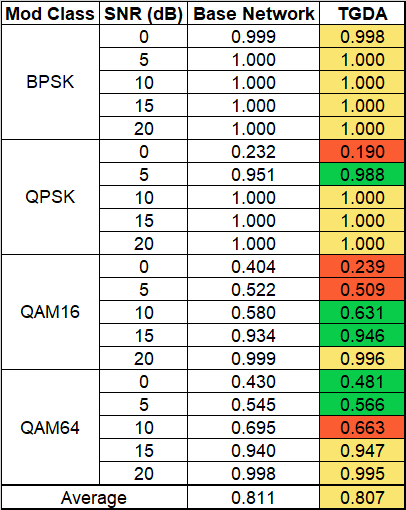}
\end{center}
\captionsetup{width = \columnwidth}
\caption{$F_1$ scores of base Model 1 network and TGDA with single component per modulation class in classifying 1000 signals at each SNR. $F_1$ within 0.01 of baseline is displayed in yellow, while higher or lower scores are green or red, respectively.  While class confusion is apportioned differently in high noise cases, average performance between the base CNN and TGDA MLE are similar.}
\label{unified_snr}
\end{figure}

While using the TGDA approach with a single component per modulation class to classify new classes was validated, its performance on the original classes was somewhat dissatisfying, though comparable to the base network, as can be seen in Figure \ref{unified_snr}. This motivated us to try a modified approach: using multiple components per modulation class.

\subsection{Modulation Classification Using SNR-Enhanced Components - Model 1}
Inputting signals with labeled SNR values into our trained Model 1 network, we visualized the Layer 7 outputs and discovered that the distributions of a modulation class's outputs were different at different SNRs. Because of this, it made sense that we could achieve better classification accuracy by breaking down each modulation class into subclasses based on SNR.

\begin{figure}[t]
\begin{center}
\includegraphics[width= \columnwidth]{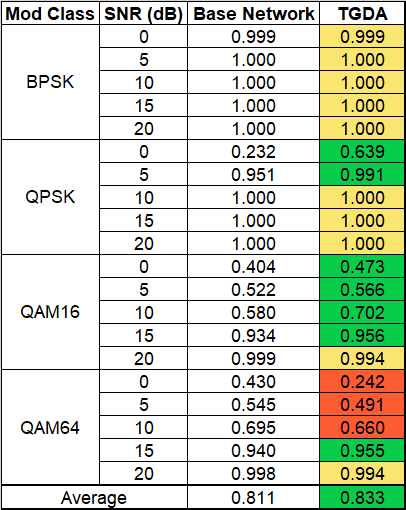}
\end{center}
\captionsetup{width = \columnwidth}
\caption{$F_1$ scores of base model 1 network vs five component per modulation class TGDA MLE in classifying 1000 signals at each SNR. $F_1$ within 0.01 of baseline is displayed in yellow, while higher or lower scores are green or red, respectively. TGDA MLE outperforms the base network on average and significantly outperforms in low-SNR QPSK. We again see different apportionment of confusion in high noise QAM signals, but almost exactly opposite the single component per modulation class case.}
\label{distinct_snr}
\end{figure}

Armed with this knowledge, we then created empirical distributions for each modulation class at 0, 5, 10, 15, and 20 dB SNR using 5,000 signal snippets for each component, and pulled feature TGDA parameters. One can view these feature distributions as conditional likelihood distributions of modulation classes given SNR. The $F_1$ scores using TGDA MLE with these components can be seen in Figure \ref{distinct_snr}.  For reference, the $F_1$ score for a random classifier in this case would be 0.25. Of note, we chose to use the maximum joint (modulation class, SNR) likelihood for classification rather than the marginal SNR likelihood to avoid biased distributions in the event that modulation classes were unequally represented across SNRs - not a concern for our controlled experiment, but potentially the case in real-world situations.

Using distinct components within each modulation class for each SNR, we see comparable or superior performance in classification relative to the base Model 1 network, with the exception of 0dB and 5dB QAM64, which each scored more than 0.05 worse than the baseline. QPSK classification performance at 0dB was significantly better with TGDA MLE than the base network, as the base network classified most of the QPSK snippets as QAM16 or QAM64.  Some insight into the TGDA performance in classifying QAM signals can be gleaned from looking at the 0dB confusion matrices in Table \ref{QAM_conf}.  Due to the way the $F_1$ score is constructed, models which divide uncertainty equally receive higher scores than those which bias towards one class. At 0dB SNR, the base network can distinguish QAM signals from PSK signals, but it's essentially a coin flip between QAM16 and QAM64.  Single component TGDA MLE has similarly poor results, with a significant bias towards QAM64, but discriminates QAM signals from other classes just as well.  Five component TGDA MLE also does a poor job distinguishing between QAM signals, with a bias towards QAM 16, and adds some confusion with QPSK signals; this is likely because the 0dB QAM signals are being compared to the 0dB QPSK distribution, rather than the less-similar average QPSK distribution across SNRs. Though there is still confusion between QAM signals at low SNR, the TDGA approach with distinct SNR-based subcomponents outperforms the base network on average and does much better at distinguishing QPSK in high-noise environments. This superior performance is expected, as we are giving our TGDA MLE method more information in the form of SNR labels on distributions.

\begin{table}[t]
\caption{Confusion Matrices for QAM Signals at 0 dB.}
\begin{center}
\renewcommand{\arraystretch}{1.2}
\renewcommand{\tabcolsep}{0.1cm}
\begin{tabular}{|l|c|c|c|c|c|c|}
\hline
 & \multicolumn{ 2}{c|}{\textbf{Base\_Network}} & \multicolumn{ 2}{c|}{\textbf{1 Comp. TGDA}} & \multicolumn{ 2}{c|}{\textbf{5 Comp. TGDA}} \\ \hline
\textbf{Mod Class} & QAM16 & QAM64 & QAM16 & QAM64 & QAM16 & QAM64 \\ \hline
BPSK & 0.000 & 0.001 & 0.000 & 0.000 & 0.000 & 0.000 \\ \hline
QPSK & 0.010 & 0.010 & 0.005 & 0.007 & 0.259 & 0.255 \\ \hline
QAM16 & \cellcolor{Gray}0.461 & 0.439 & \cellcolor{Gray}0.176 & 0.172 & \cellcolor{Gray}0.555 & 0.572 \\ \hline
QAM64 & 0.529 & \cellcolor{Gray}0.550 & 0.819 & \cellcolor{Gray}0.821 & 0.186 & \cellcolor{Gray}0.173 \\ \hline
\end{tabular}
\\[2pt] (Correct classification probabilities shaded in gray.)
\end{center}
\label{QAM_conf}
\end{table}

\subsection{Signal-to-Noise Ratio}
Despite these somewhat mixed results in modulation classification, a surprising result emerges from the data: TGDA allows us to use our trained network to estimate SNR. This ability arises because, as noted in the previous section, the feature distributions of the modulation classes are quite distinct at certain SNRs. Accuracies for signals at varying SNRs can be seen in Table \ref{SNR_conf}, with the rows of the correct SNR shaded in gray. Even when the SNR is misjudged, primarily in the 15-20 dB regime, accuracy within +/- 5dB never falls below 98\%. SNR predictions are most accurate at lower SNRs, which is when this information would be of the most benefit, as this is the region with the most confusion between modulation classes.  When taken in conjunction with the modulation class prediction's logarithmic likelihood, knowledge of the SNR value allows us to further calibrate our confidence in the predicted class. Knowing the conditional probabilities of a modulation class given a predicted class and SNR is of value in a distributed sensor environment, as it allows us to weight the predictions of sensors which receive higher-SNR signals, rather than treat all sensor predictions equally.

\begin{table}[t]
\caption{SNR Confusion Matrix for Model 1-Based TGDA. Zero-confusion values are not shown.}
\begin{center}
\renewcommand{\arraystretch}{1.2}
\renewcommand{\tabcolsep}{0.1cm}
\begin{tabular}{|l|r|r|r|r|r|}
\hline
\textbf{} & \multicolumn{1}{l|}{\textbf{}} & \multicolumn{ 4}{c|}{\textbf{Mod Class}} \\ \hline
\textbf{Actual SNR} & \multicolumn{1}{l|}{\textbf{Predicted SNR}} & \multicolumn{1}{l|}{\textbf{BPSK}} & \multicolumn{1}{l|}{\textbf{QPSK}} & \multicolumn{1}{l|}{\textbf{QAM16}} & \multicolumn{1}{l|}{\textbf{QAM64}} \\ \hline
\multicolumn{ 1}{|c|}{0} & 0 & \cellcolor{Gray}1 & \cellcolor{Gray}0.999 & \cellcolor{Gray}0.997 & \cellcolor{Gray}0.999 \\ \cline{ 2- 6}
\multicolumn{ 1}{|c|}{} & 5 & 0 & 0.001 & 0.003 & 0.001 \\ \hline
\multicolumn{ 1}{|c|}{5} & 0 & 0 & 0.004 & 0.002 & 0 \\ \cline{ 2- 6}
\multicolumn{ 1}{|l|}{} & 5 & \cellcolor{Gray}0.961 & \cellcolor{Gray}0.979 & \cellcolor{Gray}0.989 & \cellcolor{Gray}0.996 \\ \cline{ 2- 6}
\multicolumn{ 1}{|l|}{} & 10 & 0.037 & 0.017 & 0.009 & 0.004 \\ \cline{ 2- 6}
\multicolumn{ 1}{|l|}{} & 15 & 0.002 & 0 & 0 & 0 \\ \hline
\multicolumn{ 1}{|c|}{10} & 5 & 0.12 & 0.007 & 0.004 & 0.001 \\ \cline{ 2- 6}
\multicolumn{ 1}{|l|}{} & 10 & \cellcolor{Gray}0.822 & \cellcolor{Gray}0.936 & \cellcolor{Gray}0.939 & \cellcolor{Gray}0.846 \\ \cline{ 2- 6}
\multicolumn{ 1}{|l|}{} & 15 & 0.058 & 0.057 & 0.057 & 0.138 \\ \cline{ 2- 6}
\multicolumn{ 1}{|l|}{} & 20 & 0 & 0 & 0 & 0.015 \\ \hline
\multicolumn{ 1}{|c|}{15} & 10 & 0.054 & 0.113 & 0.064 & 0.108 \\ \cline{ 2- 6}
\multicolumn{ 1}{|l|}{} & 15 & \cellcolor{Gray}0.817 & \cellcolor{Gray}0.771 & \cellcolor{Gray}0.739 & \cellcolor{Gray}0.468 \\ \cline{ 2- 6}
\multicolumn{ 1}{|l|}{} & 20 & 0.129 & 0.116 & 0.197 & 0.424 \\ \hline
\multicolumn{ 1}{|c|}{20} & 10 & 0 & 0.001 & 0 & 0.012 \\ \cline{ 2- 6}
\multicolumn{ 1}{|l|}{} & 15 & 0.326 & 0.172 & 0.272 & 0.197 \\ \cline{ 2- 6}
\multicolumn{ 1}{|l|}{} & 20 & \cellcolor{Gray}0.674 & \cellcolor{Gray}0.827 & \cellcolor{Gray}0.728 & \cellcolor{Gray}0.791 \\ \hline
\end{tabular}
\\[2pt] (Correct classification proportions shaded in gray.)
\end{center}
\label{SNR_conf}
\end{table}

\subsection{Samples Per Symbol}\label{4class SPS}
\begin{figure}[t]
\begin{center}
\includegraphics[width= \columnwidth]{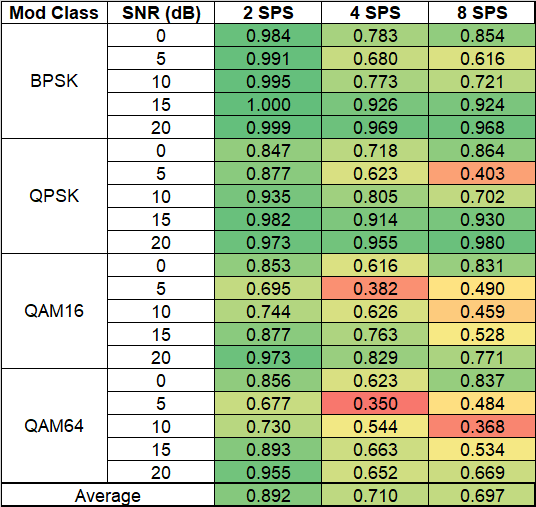}
\end{center}
\captionsetup{width = \columnwidth}
\caption{$F_1$ Scores of Model 1-Based TGDA MLE in Estimating Samples Per Symbol (SPS).}
\label{sps_F1}
\end{figure}

Next, we applied GDA MLE to the problem of predicting samples per symbol (SPS) of a given signal.  This is a somewhat different problem from the SNR prediction, as the base network was only trained on 4 SPS signal snippets and logically would not have included SPS discrimination as a useful ability for classification.  $F_1$ scores of this distribution-based approach can be seen in Table \ref{sps_F1}. For reference, the $F_1$ score for a random classifier in this case would be 0.333.  None of the 75 values in this table fall below the random threshold, and 58 are more than twice this value.  Performance is generally better at higher SNRs and fewer samples per symbol, with few exceptions; intuitively, a clearer signal (high SNR) and more observed symbols (low SPS) would provide the classifier with better quality information.

\section{Dissimilar Modulation Classification and Network Comparison Experiments}\label{Comparative Experiments}
The results from our first set of experiments raised some interesting questions, which we explore here.  For example, we demonstrated that a phase- and amplitude-modulated signal class could be learned using a network trained on other phase- and amplitude-modulated signal classes; here we assess the performance of TGDA MLE with new modulation types, e.g. frequency shift keyed (FSK) signals. We also assess how changing the type of training data affects subsequent TGDA MLE effectiveness, with regard to modulation, SNR, and SPS classification. Finally, we compare the performance of TGDA MLE based on more narrowly-trained networks to that of a convolutional neural network trained on all classes of interest at varying signal-to-noise ratios and samples per symbol.

\subsection{Performance vs FSK Signals}
\begin{figure*}[ht]
\centerline{\includegraphics[width=\textwidth]{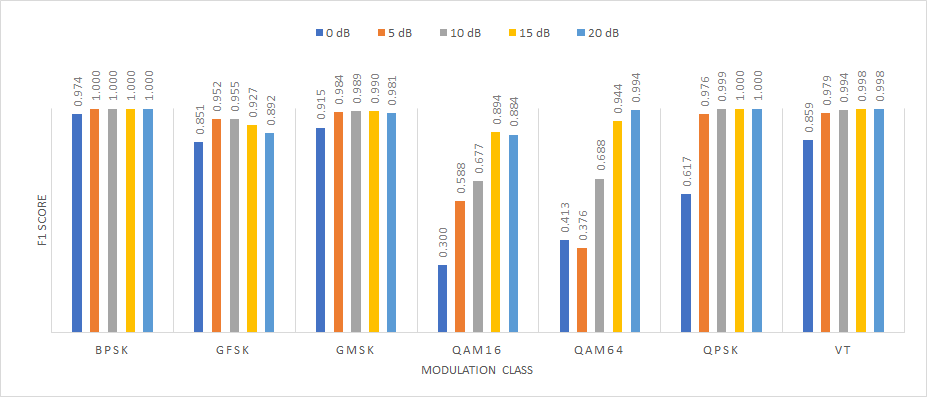}}
\caption{F1 scores for classification of seven modulation classes at 4 SPS using Model 1-based TGDA MLE.}
\label{Model1_7class}
\end{figure*}

To stress the utility of TGDA MLE in classifying new modulation classes, we added Gaussian frequency shift keying (GFSK) and Gaussian minimum shift keying (GMSK) signal types to the test data. Parameters were extracted for these new classes in the same manner as previously, and we attempted classification using an equal mix of each of the seven signals at 0, 5, 10, 15, and 20dB SNR.  All signals in this case were sampled at 4 SPS.  Results can be seen in Figure \ref{Model1_7class}. Despite having no FSK-type signals in the data on which the CNN was trained, the network has clearly learned features that not only allow it to identify FSK-type signals but also to distinguish between generic GFSK signals and GMSK signals. However, note the degradation in performance of GFSK at 15 and 20dB SNR, and a similar reduction in performance of QAM16 at the same SNRs relative to our four- and five-class scenarios: there is actually significant confusion between QAM16 and GFSK at these SNRs. We see this as a cautionary tale: if the types of features which characterize new modulation classes are not seen in the CNN's training data, there is a greater risk for confusion when using the CNN's feature space to classify these modulation classes.

\subsection{Modulation Classification Comparison}
To see how variations in the types of data used to train the CNN can affect TGDA MLE based on this CNN, we trained four other networks as described in Table \ref{tab:training_networks} and repeated our experiments using these new networks.  We also tested against signals sampled at 2 and 8 samples per symbol. 

\begin{figure}[t]
\begin{center}
\includegraphics[width= \columnwidth]{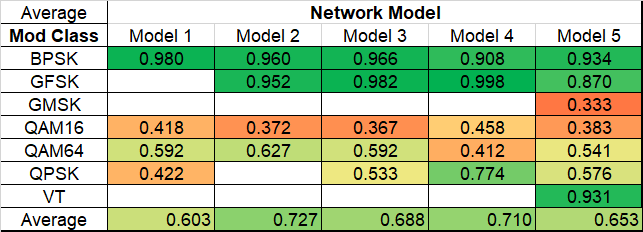}
\end{center}
\captionsetup{width = \columnwidth}
\caption{Average $F_1$ scores for CNN classification of seven modulation classes at 2, 4, and 8 SPS and 0-20dB SNR for each model.}
\label{Net_Comparison}
\end{figure}

For a baseline comparison, $F_1$ scores for the base networks in classifying signals at 2, 4, and 8 SPS from their respective training modulation classes are shown in Figure \ref{Net_Comparison}. Because Models 1, 2, and 3 were only trained on 4 SPS signals, we would expect their performance in this evaluation to suffer relative to Models 4 and 5, which were trained on 2, 4, and 8 SPS signals. While this was generally true for the 2 SPS signals, Models 1, 2, and 3 actually performed comparably to the better-trained models in classifying 8 SPS signals. Note that Model 5's performance in classifying GFSK is lower than other models - interestingly, this is because of confusion with GMSK (not present in other models), but only at 2 and 8 SPS. It is unclear why Model 5 learned to distinguish these classes at only 4 SPS when it was trained on all SPS levels.

\begin{figure}[ht]
\includegraphics[width= \columnwidth]{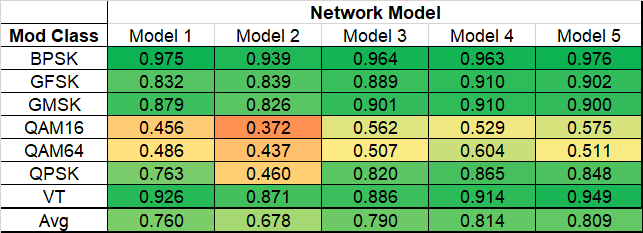}
\captionsetup{width=\columnwidth}
\caption{$F_1$ scores for TGDA MLE classification of seven modulation classes at 2, 4, and 8 SPS and 0-20dB SNR for each model.}
\label{Dist_Comparison}
\end{figure}

The TGDA MLE results are shown in Figure \ref{Dist_Comparison}. Differences in average $F_1$ scores of TGDA MLE and the base CNNs in classifying training modulation classes are shown in Table \ref{Distr_Delta}. Note that the comparison is between the CNNs classifying the original classes and TGDA MLE classifying all classes, so the TGDA MLE data is subjected to additional opportunities for confusion. From this data, interesting trends emerge. 

First, and most obviously, training data matters. It is apparent from these results that the presence of QPSK in the training data is critical for learning the features which distinguish signals with quadripolar / quadrature characteristics, as TGDA MLE using the model without QPSK in the training data, Model 2, performs considerably worse in these classes. Average performance also increases in the models with more training classes (Models 3-5) relative to those with fewer, though this performance improvement is constrained by using the same architecture in all models (i.e. at some point adding training classes will saturate the CNN's capacity for learning discriminant features).

Second, while we would expect performance across the board to suffer slightly due to confusion with added classes, this is not always the case. For example, while almost the entire difference in BPSK performance in Model 1 and 2 TGDA MLE is the result of confusion with the added VT signal at 0 and 5 dB SNR, QPSK classification in all models improves significantly using TGDA MLE, despite the added classes. 

Third, when we control for the addition of new signal types, the performance improvements from TGDA MLE are notable: GFSK classification performance diminishes in Models 2, 3, and 4 due to confusion with the GMSK class, which wasn't present in the training data; however, the improvement in Model 5 TGDA MLE in distinguishing these classes is significant. In fact, while there is a slight reduction in classification of QAM64 relative to the base CNN, Model 5 TGDA MLE otherwise shows improvement across all classes. 

Fourth, compare the average $F_1$ scores of Model 5 in figure \ref{Net_Comparison} to the values of all models in Figure \ref{Dist_Comparison}. Even though Model 2's TGDA MLE performance is substandard in QAM and QPSK, the average $F_1$ score across all modulation classes still exceeds that seen for the base CNN of Model 5. Other models fare even better, demonstrating the effectiveness of TGDA MLE in modulation classification.

While not reflected in the figures, our results also echoed the observation made in \cite{karra_modulation_2017} regarding the favorability of longer input snippets. Across all models tested, the $F_1$ scores of the 8 SPS signals were almost always significantly worse than those of the 2 and 4 SPS signals. Because all snippets comprised the same number of samples, higher SPS signals effectively gave the network fewer symbols to analyze, effectively shortening the ``snapshot" and increasing the potential for confusion.

\begin{table}[t]
\begin{center}
\renewcommand{\arraystretch}{1.2}
\begin{tabular}{|c|c|c|c|c|}
\hline
Model 1 & Model 2 & Model 3 & Model 4 & Model 5 \\ \hline
0.067 & -0.081 & 0.060 & 0.064 & 0.156 \\ \hline
\end{tabular}
\end{center}
\caption{Difference in average $F_1$ scores of TGDA MLE relative to CNN classification of training classes. Positive values indicate superior performance of TGDA MLE. TGDA MLE scores are from the seven modulation class problem in all cases, while CNN scores are from only the training classes of each model.}
\label{Distr_Delta}
\end{table}

\subsection{SNR Estimation Comparison}
\begin{figure}[t]
\includegraphics[width= \columnwidth]{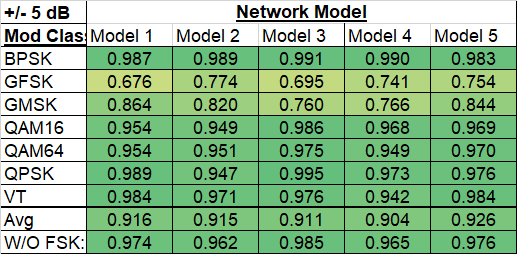}
\captionsetup{width=\columnwidth}
\caption{+/- 5dB SNR estimation accuracy for TGDA MLE classification by class at 2, 4, and 8 SPS.}
\label{SNR_win5}
\end{figure}

\begin{figure}[t]
\includegraphics[width= \columnwidth]{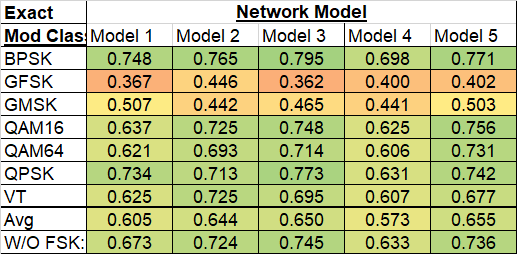}
\captionsetup{width=\columnwidth}
\caption{SNR estimation accuracy for TGDA MLE classification by class at 2, 4, and 8 SPS.}
\label{SNR_exact}
\end{figure}

Next we examined the effect of training data on SNR estimation. As before, we use accuracy instead of $F_1$ scores, first showing accuracy within 5dB SNR in Figure \ref{SNR_win5}, then looking for an exact match (at 5dB intervals) in Figure \ref{SNR_exact}. We see that in both the +/- 5dB and exact cases, adding modulation classes to, and varying the SPS rate of, the training data does not have a significant effect on SNR estimation. We also see that across all models, the ability to predict the SNR of FSK signals is poor relative to other modulation types. Even when both GFSK and GMSK are present in the training data (thus necessitating differentiating between the signals in the CNN), it seems the network does not identify SNR as a useful characteristic in discriminating between these modulation classes. Clearly, the mere presence of information in training data is not sufficient to ensure that this information will be encoded as a feature. However, when we remove the GMSK and GFSK results, we see that +/- 5dB SNR estimation accuracy averages above 96\% across each model, and the exact SNR is estimated correctly at least 63\% of the time. As we saw with the Model 1 four- and five-class experiments, SNR estimation accuracy is highest at lower SNRs. In fact, disregarding the FSK classes, the +/-5 dB SNR accuracy at 0 and 5 dB was at least 94.3\% for all models tested at all samples per symbol.

When we compare SNR results among the various models, few clear trends emerge. Simply including more classes in test data does not ensure better results: though Model 5's +/- 5dB accuracy is marginally higher than the other models, Model 4's is lower than that of Model 3 (trained with the same classes but only 4 SPS) and even that of Model 1 (trained with only 4 SPS and without GFSK). Seeing GFSK in training seems to improve SNR accuracy for that modulation class, but despite no FSK exposure in training, Model 1's GMSK noise estimation exceeded all other models tested.

\subsection{SPS Estimation Comparison}
\begin{figure}[ht]
\includegraphics[width= \columnwidth]{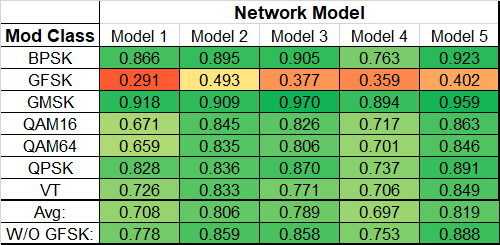}
\captionsetup{width=\columnwidth}
\caption{Average $F_1$ scores in classifying samples per symbol by model and modulation class at 0-20dB SNR.}
\label{SPS_F1_fig}
\end{figure}

Noting the success of using TGDA MLE to estimate SNR, we next sought to see how our five trained models performed in classifying samples per symbol. As in the four-class problem discussed in subsection \ref{4class SPS}, we tested TGDA MLE's ability to discriminate between 2, 4, and 8 SPS signals; however, we did this for all models and all seven test classes. Average results by model and class are displayed in Figure \ref{SPS_F1_fig}. 

\begin{figure}[t]
\includegraphics[width= \columnwidth]{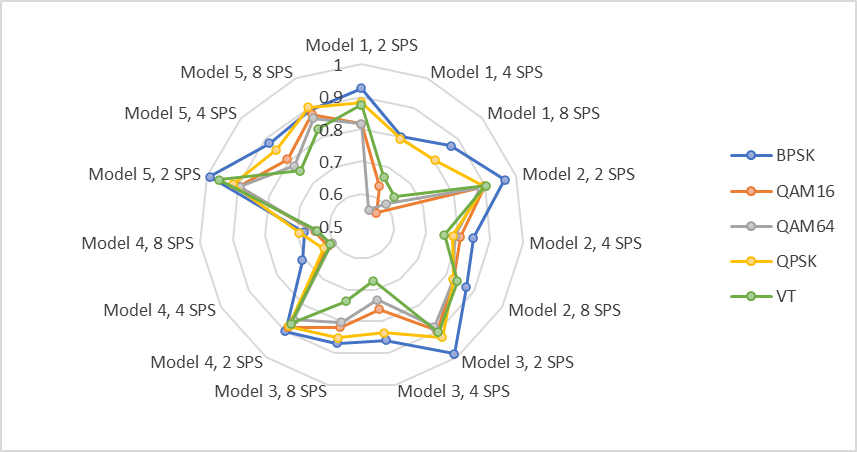}
\captionsetup{width=\columnwidth}
\caption{$F_1$ scores in classifying samples per symbol by SPS, model, and modulation class at 0-20dB SNR. TGDA MLE for all models is generally good at identifying 2 SPS signals but has a harder time differentiating between 4 and 8 sps signals.}
\label{SPS_F1_Radar}
\end{figure}

While results by SNR are not shown, SPS classification performance was lowest at 0dB SNR and increased with SNR - so as with modulation classification, confidence in SPS classification results can be calibrated with SNR estimation. Comparisons of models based on neural network training data are inconclusive: Model 5 performs the best, as we would expect, given its neural network was trained on all classes and SPS; however, Model 4, which was also trained on all samples per symbol, performed worse than Models 1-3, which were trained on narrower data sets. It is not clear why SPS classification of GFSK signals was so consistently poor, as these signals' physical characteristics have much in common with GMSK. Much like the four-class problem, we saw significantly better performance in classifying 2 SPS signals (greater than 0.85, with the exception of GFSK) than 4 or 8 SPS signals across all models. Results by SPS are displayed in Figure \ref{SPS_F1_Radar}.

\section{Conclusion}
\subsection{Summary}
By using a distribution-based classification approach, we are able to share information about new classes among sensors by sending only 200 parameters per class, whereas sending a retrained CNN would require 8,551,642 parameters (the retrained trainable parameters from the original network), plus 50 additional parameters (the final dense layer's weights) for each added class - assuming no additional convolutional filters or features are added to the retrained network. This advantage can be obtained while enjoying classification accuracies comparable to - and in many cases better than - those of a CNN trained on the new class. We also saw that information about physical features of interest - SNR and SPS - was captured in the learned features, despite not explicitly training the CNN to discriminate between these features.

Ultimately, there are two primary takeaways of this work. First, TGDA MLE, or a comparable distribution-based approach, can be used to learn classes similar to those of the training data without retraining the network - saving time, reducing hardware requirements, and supporting a low-bandwidth, low-power sensor network. As more information becomes available about training signals (such as SNR labels), more detailed subclasses and conditional probability distributions can be formed, providing more accurate classification along with measures of confidence in classification results. Second, as we saw with SNR estimation of FSK signals in Model 2, the mere presence of information in training data is no guarantee of seeing it encoded as meaningful features, so any ability to discriminate features of interest beyond modulation classification using the current TGDA MLE approach should be seen as a potential benefit rather than a primary capability. The best way to utilize distribution-based approaches may be to have the CNN learn relevant physical features, rather than classes, then use distributions across these features to define the classes, much as a human expert would do. Such an approach would allow classification, meaningful description of features, and generation of hypothetical signals using new combinations of known features.

\subsection{Future Work}
The results in this paper were achieved without any dimensionality reduction or feature weighting.  It would be useful to see if these results could be replicated or improved upon after applying principal components analysis or feature selection to the 50 nodes of Layer 7. In addition to speeding inference, such processing might identify and quantify correlation of nodes to specific attributes (such as SNR), increasing accuracy and allowing regression of features for classification of attributes with values outside the training set.

A key element of the TGDA MLE process is the convenient distribution of signal features as mixtures of truncated Gaussians, enabling characterization of classes by a small number of parameters. While all models tested in this work presented features of this type, success of TGDA MLE in other contexts requires determining which network architectures or elements result in these types of feature distributions, or at the very least that these distributions are present before continuing with MLE.

TGDA MLE provides likelihood values for each feature, which, while used in this work for classification, might also be used to identify outlier signals. However, we have not validated novel signal detection and characterization using TGDA MLE. Identifying and characterizing new signals in the field using this approach would first require thresholding feature likelihood values, so research should be conducted into robust methods for selecting these thresholds.

Though the data generated for this paper was simulated, it would be interesting to assess the performance of TGDA MLE on real-world signals.  Using signals from different emitters, one might determine specific emitter effects on signal feature distributions.  While training the network on and classifying physical signals would be a validation of the approach generally, successful inference of physical signals with a system trained on simulated signals (perhaps with the addition of frequency offset and sample mismatch) would allow us to anticipate real-world signals of interest by creating a greater variety of prior feature distributions.

Finally, this paper looked at additional modulation classes, SNR, and samples per symbol; what other signal attributes might be inferred using this approach?  The success of maximum likelihood estimation in estimating SNR may be a result of training the original network on a range of SNRs; if we trained the base network on a narrower range of SNRs, or even a single SNR, would the trained network still be able to discriminate this attribute?  If we could qualify what physical features are represented in the trained network based on the data used to train it, we could ensure that future training sets are sufficiently diverse to allow the network to capture all features of interest. Additionally, with high confidence of which features might be learned from a training set, we could minimize the training samples required to learn the necessary features, thus reducing training time and data burden.


%

\section*{Acknowledgment}
This research was funded by Office of Naval Research fundamental research grant N00014-17-1-2835.  Any opinions, findings, and conclusions or recommendations expressed in this material are those of the authors and do not necessarily reflect the views of the Office of Naval Research.

The authors would like to thank Seth Hitefield, who helped clean up our Python code, navigate GNU Radio, and significantly speed up the data generation process, and to Parker White, who provided the original network training architecture.



\bibliographystyle{IEEEtran}
\bibliography{main_paper}


%

\begin{IEEEbiography}[{\includegraphics[width=1in, height=1.25in, trim=30 0 30 0, clip, keepaspectratio]{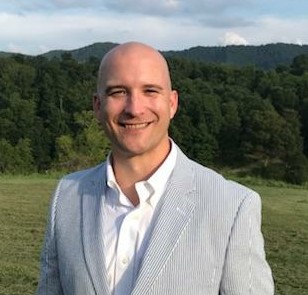}}]{J.B. Persons}
is a direct-PhD student in Computer Engineering and a graduate research assistant with the Ted and Karyn Hume Center for National Security and Technology at Virginia Tech. His research interests in machine intelligence include edge and low-data learning systems as well as command and control architectures for distributed systems. J.B. received a B.S. in Mechanical Engineering from the Massachusetts Institute of Technology and an M.S. in Computer Engineering from Virginia Tech, and he possesses tactical aviation, S\&T program management, and operational experience from his time in the U.S. Marine Corps.
\end{IEEEbiography}

\begin{IEEEbiography}[{\includegraphics[width=1in,height=1.25in,trim=20 0 20 0,clip,keepaspectratio]{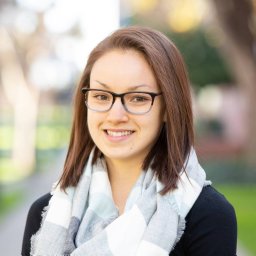}}]{Lauren J. Wong}
is a Deep Learning Data Scientist with Intel Corporation, and was a research associate at the Hume Center for National Security and Technology at Virginia Tech at the time of this work. She received the M.S. degree in electrical engineering from Virginia Tech, and the B.A. degrees in computer science and mathematics from Oberlin College.
\end{IEEEbiography}

\newpage

\begin{IEEEbiography}[{\includegraphics[width=1in,height=1.25in,clip,keepaspectratio]{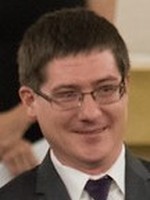}}]{W. Chris Headley}
Dr. William ``Chris" Headley is the Associate Director for the Electronic Systems Laboratory within the Ted and Karyn Hume Center for National Security and Technology at Virginia Tech where he has served as a principal or co-principal investigator on a multitude of government and commercial projects totaling over \$11M. Within the lab he oversees the Radio Frequency Machine Learning (RFML) portfolio. Through his courtesy appointment within Virginia Tech’s Electrical and Computer Engineering department, he also serves as a mentor and advisor to both undergraduate and graduate student researchers, providing them with hands-on research opportunities through these projects as well as guiding them towards their degree requirements. Dr. Headley earned his BS/MS/PhD in Electrical Engineering at Virginia Tech. He has written over 20 conference/journal publications. His current research interests include spectrum sensing, radio frequency machine learning, and virtual reality educational opportunities.
\end{IEEEbiography}

\begin{IEEEbiography}[{\includegraphics[width=1in,height=1.25in,clip,keepaspectratio]{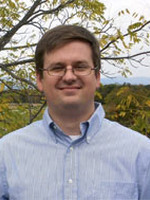}}]{Michael C. Fowler}
Dr. Michael Fowler is the Assistant Director for Autonomous and Multiagent Systems for the Ted and Karyn Hume Center for National Security and Technology at Virginia Tech responsible for driving and providing thought leadership into the center's research on autonomous systems, mission orchestration, distributed intelligence, and security for wireless and unmanned systems. 
His research focus is on the convergence of distributed intelligence and machine learning for decision making under uncertainty for embedded applications including drones, satellites, IoT, and wireless communications. 
He has accumulated over 10 years of experience managing and performing research in security, wireless systems and artificial intelligence at Harris Corporation and as faculty at Virginia Tech. 
He has received a Masters in Engineering Management from Old Dominion University and a Ph.D. in Computer Engineering from Virginia Tech. 
\end{IEEEbiography}

\end{document}